\documentclass[12pt]{iopart}
\usepackage{graphicx}
\begin{document}
\jl{1}
\title{The specific heat of the two-dimensional $\pm J$ Ising
model}

\author{Hidetsugu 
Kitatani\footnote[3]{E-mail address:kitatani@vos.nagaokaut.ac.jp},
Toshimasa Chino and Hideshi Ohya}

\address{Department of Electrical Engineering, Nagaoka University of
Technology, Nagaoka, Niigata 940-2188, Japan}

\begin{abstract}
The specific heat of the two-dimensional $\pm J$ Ising model has been
investigated by the numerical transfer matrix method and Monte Carlo
simulations from a new point of view.
The region where a part of the specific heat 
takes the negative value has been investigated,
which is characteristic of
frustrated systems and reflects the non-trivial degeneracy of the ground state.
The region mentioned above is found to be fairly
large in the $p-T$ plane
($p$ is the concentration of the ferromagnetic bond and $T$ is
the temperature). Moreover, it includes the Nishimori line.
Namely, it includes a part of the 
paramagnetic-ferromagnetic phase boundary,
on which the specific heat cannot diverge.
The present result indicates that
the specific heat does not diverge at least on a part of the
paramagnetic-ferromagnetic phase boundary
above the multicritical point, which is in conflict with
previous results. 
\end{abstract}

%
%
\pacs{75.50.Lk,02.70.Lq,64.60.Cn,05.50.+q}

\section{Introduction}
To elucidate the nature  of critical phenomena
of two-dimensional disordered Ising models
has been a subject of a long-standing interest.  Harris[1] concluded
from a heuristic argument that the nature of critical phenomena of  a
disordered
system becomes different from that of the corresponding pure system
when
 the critical exponent of the 
specific heat, $\alpha$, of the pure system is positive, while it remains
the same when $\alpha<0$.
Since the two-dimensional pure ferromagnetic Ising model
is the marginal case, namely $\alpha =0$,
 many authors have
investigated 
the properties of critical phenomena of
the two-dimensional disordered Ising models[2-14].

For the two-dimensional unfrustrated random Ising models,
many authors concluded that the specific heat diverges
double logarithmically
 at the paramagnetic-ferromagnetic phase boundary[2-9], 
though there are several results which
insist that the specific heat remains finite[1,10-12].

For the two-dimensional $\pm J$ Ising model which corresponds
to the two-dimensional frustrated random Ising model, a few results exist.
Inoue[13] investigated the
system-size dependence of the height of the peak of the specific heat
at $p\geq 0.94$, and
concluded that the logarithmic divergence of the specific heat
at the paramagnetic-ferromagnetic phase boundary is most probable
in the same way as the pure case. Reis {\it et al}[14]  investigated 
the system-size dependence of the correlation length at $p\geq 0.92$, and 
concluded that logarithmic corrections do not play a role 
in contrast with the unfrustrated disordered systems.
They concluded, however,
that the specific heat diverges at most logarithmically. Namely,
they could not completely 
exclude the possibility of
the double logarithmic divergence of the specific heat.

On the other hand, there exisits the Nishimori line,
which is defined by the equation, $\exp (2J/k_{\rm B}T)=p/(1-p)$.
($k_{\rm B}$ is the Boltzmann constant, which we put that $k_{\rm B}=1$
from now on.) On the Nishimori line, several rigorous results 
have been derived[15]. Particularly, it has been proved that the specific
heat remains finite.

From the renormalization group approach, it is generally
believed that, for the two-dimensional $\pm J$ Ising model,
there exist two fixed points on the paramagnetic-ferromagnetid
phase boundary above the multicritical point, 
namely the pure ferromagnetic fixed point and the multicritical 
fixed point[16-17],
and there seems 
to be no random fixed point in contrast with the three-dimensional case[18].
(In this paper, we use the word "multicritical point" as the 
crossing point of the ferromagnetic-nonferromagnetic phase boundary
and the Nishimori line, though there seems to be no spin glass phase
in the two-dimensional case.)
The pure ferromagnetic fixed point is stable, and the multicritical fixed
point is unstable.
Thus, the critical
phenomena on the paramagnetic-ferromagnetic phase boundary above the
multicritical point is governed by the pure ferromagnetic fixed point.
Therefore, the critical exponent, $\alpha$, should be zero on the
whole paramagnetic-ferromagnetic phase boundary above the multicritical point.
The fact mentioned above, however, does not make a help to determine
the critical behaviour of the specific heat, since each
of
a logarithmic divergence,
a double logarithmic divergence and a cusp like behaviour belongs to the
case, $\alpha =0$.

 In this paper, we investigate the specific heat 
 from a
new point of view.
We devide the specific heat of a system into two parts, $C_{1}$
and $C_{2}$.
(For the detail definitions of $C_{1}$ and $C_{2}$, see section 2.)
The value of $C_{1}$ is easily found to be non-negative
and remain finite at finite temperature. 
On the other hand,  the value of $C_{2}$
 becomes negative at $T=0$ when the system is frustrated and the ground state
 has non-trivial
degeneracy.
Since
we can consider that the negative value of
 $C_{2}$  reflects
the non-trivial ground state degeneracy,
it is an interesting problem to make it clear up to
 what temperature the property persists.
It is noted that in order that the specific heat may diverge, $C_{2}$
should become infinite.

Thus, we investigate the region in the $p-T$ plane
where $C_{2}$ takes the negative value
for the two-dimensional $\pm J$ Ising model.
Our result shows that the region mentioned above
is fairly large in the $p-T$ plane. Moreover, it includes
 the Nishimori line.
Namely, the region includes a part of the 
paramagnetic-ferromagnetic phase boundary above the 
multicritical point.
In that region, the specific heat cannot diverge.

The present result indicates that the specific heat does not diverge
at least on a part of the paramagnetic-ferromagnetic
phase boundary 
near and above the multicritical point.  The result is not 
directly in conflict  with previous results[13,14] which
insist that
the specific heat diverges at most logarithmically
at the paramagnetic-ferromagnetic phase boundary
near the pure ferromagnetic case($p \geq 0.92$), since the range of 
the concentration
of the calcultions does not overlap.
It
is natural, however,  to think that the nature of the phase transition
does not change on the paramagnetic-ferromagnetic phase
boundary above the multicritical point. Thus,  we insist that
the specific heat of the two-dimensional $\pm J$ Ising model
does not diverge on the whole paramagnetic-ferromagnetic phase
boundary above the multicritical point, though we cannot exclude the possibility  that there exists a singular point on the paramagnetic-ferromagnetic
phase boundary which separates the nature of the specific heat.

\section{The model and the method}
We consider the two-dimensional $\pm J$ Ising model on 
a square lattice with only nearest neighbor interactions.
The Hamiltonian  is written as follows:
\begin{equation}
 {\cal H} = -\sum_{(ij)}J_{ij}S_{i}S_{j}, 
\end{equation}
where $S_{i}=\pm 1$, and the
summation of $(ij)$ runs over all the nearest neighbors.
 Each $J_{ij}$ is determined according to the following
probability distribution:
\begin{equation}
    P(J_{ij}) = p\delta (J_{ij}-J)+(1-p)\delta (J_{ij}+J).
\end{equation}
In this paper, we put that $J=1$.

The specific heat per bond, $C(p,T)$, is written as follows;
\begin{equation}
   C(p,T)=\frac {1}{N_{\rm B}}\frac{\partial }{\partial T}
[<E>_{T}]_{p} =\frac {1}{N_{\rm B}}\frac{\partial }{\partial T}
\sum_{(ij)}[<-J_{ij}S_{i}S_{j}>_{T}]_{p},
\end{equation}
where $<\cdots >_{T}$ denotes the thermal average
in a given bond configuration, $\{J_{ij} \}$, at temperature, $T$.
$[\cdots ]_{p}$ denotes the configurational average at the
ferromagnetic bond concentration, $p$,
and $N_{\rm B}$ is the number of bonds.

Now, we divide the specific heat of the system, $C(p,T)$ into two parts,
$C_{1}(p,T)$ and $C_{2}(p,T)$:
\begin{equation}
 C(p,T)=C_{1}(p,T)+C_{2}(p,T),
\end{equation}
where
\begin{equation}
   C_{1}(p,T)=\frac {1}{N_{\rm B}}
    \sum_{(ij)}\frac{\partial }{\partial T_{ij}}
  [<-J_{ij}\sigma_{i}\sigma_{j}>_{\{T\}}]_{p}\mid_{\{T\}=T},
\end{equation}
and
\begin{equation}
  C_{2}(p,T)=\frac {1}{N_{\rm B}}\sum_{(ij)}
\sum_{lm \neq ij} \frac {\partial}{\partial T_{lm}}[<-J_{ij}\sigma_{i}\sigma_{j}>_{\{T\}}]_{p}\mid_{\{T\}=T}.
\end{equation}
Here,  we introduce technically, $T_{ij}$, the local temperature
of the bond, $J_{ij}$.
Namely, $C_{1}(p,T)$ is considered to be the 
configurational average of the change of the local energy, $-<J_{ij}\sigma_{i}\sigma_{j}>_{\{T\}}$, when we infinitesimally
increase the correspondeing local temperature, $T_{ij}$.
On the other hand, $C_{2}(p,T)$ is considered to be the 
configuratinal average of the the change of the local energy, $-<J_{ij}\sigma_{i}\sigma_{j}>_{\{T\}}$, when we infinitesimally
increase the temperarture, $T_{lm}$, which surrounds the local bond, $J_{ij}$.

It is easily calculated that
\begin{equation}
   C_{1}(p,T)= \frac {1}{N_{\rm B}T^{2}}\sum_{(ij)}(1 -  [<\sigma_{i}\sigma_{j}>_{T}^{2}]_{p}).
\end{equation}
Namely, $C_{1}(p,T)$ has an upper bound;
\begin{equation}
    C_{1}(p,T) \leq  \frac {1}{T^{2}}.
\end{equation}
Therefore, at finite temperature, in order that the specific heat may diverge, 
$C_{2}(p,T)$ should become infinite.

Now, we consider the case at zero temperature.
At zero temperature, the specific heat, $C(p,0)$ becomes zero.
When, there is no non-trivial degeneracy in the ground state of the
system, each $[<\sigma_{i}\sigma_{j}>_{T}^{2}]_{p}=1$.
Namely, $C_{1}(p,0)$ becomes zero. Thus, $C_{2}(p,0)$ also becomes zero.
On the other hand, when the ground state has non-trivial degeneracy,
some of $[<\sigma_{i}\sigma_{j}>_{T}^{2}]_{p}$ become less than one.
Namely, $C_{1}(p,0)$ becomes positive infinite. 
Thus, $C_{2}(p,0)$  becomes negative infinite in this case.
The negative value of $C_{2}(p,T)$ at finite temperature
may be considered to be
one of the influence of the non-trivial ground state degeneracy of the
frustrated system.  Thus, it is an intersting problem 
to make it clear up to what temperature
the property persists.
Therefore, in the following sections, we investigate the region
 in the $p-T$ plane where $C_{2}(p,T)$ takes  the negative value
for the two-dimensional $\pm J$ Ising model.

\section{Results by the numerical transfer matrix method}
In this section, we investigate the region in
the $p-T$ plane where $C_{2}(p,T)$ takes the
negative value by the numerical transfer matrix method.
We have calculated for the lattice size, $L=4-16$, at
the bond concentration, $p=0.65-0.95$.  In the calculations, 
we take the free
boundary condition, and 
take the configurational average for $10^{4}-10^{5}$ bond configurations.

Figure 1 shows the temperature dependence of $C(p,T)$, $C_{1}(p,T)$
and $C_{2}(p,T)$ for $L=4$ at $p=0.8$.  We can see that $C_{2}(p,T)$
takes the negative value in the low temperature region.
We have estimated the temperature, $T_{0}(L,p)$, where $C_{2}(p,T)$ takes
the value zero for $L=4-16$ at $p=0.65-0.95$.
Figure 2 shows the temperature dependence of $C_{2}(p,T)$ near the zero point
for $L=14$ at $p=0.9$. From the figure, we have estimated
the temperture of the zero point, 
$T_{0}(L=14,p=0.9)=1.2764(14)$.
The estimated values of $T_{0}(L,p)$ for various $L$ and $p$ 
are shown in table 1. The accuracy of the values of $T_{0}(L,p)$ becomes worse
as the concentration, $p$, becomes small, since the gradient of $C_{2}(p,T)$
near the zero point becomes small.

We can see that, at each concentration, $p$, the size dependence of
the values of $T_{0}(L,p)$ is very small, and the values are fairly large
compared to
the value of $T_{\rm N}(p)$, the tempearature of the Nishimori line at
$p$. 

\begin{table}
\caption{The estimated value of $T_{0}(L,p)$. $T_{\rm N}(p)$ is the
temperature of the Nishimori line at $p$.}
\begin{indented}
\item[]\begin{tabular}{@{}llllllll}
\br
$L$&$p=0.65$&$p=0.7$&$p=0.75$&$p=0.8$&$p=0.85$&$p=0.9$&$p=0.95$\\
\mr
4&3.88(6)&2.85(5)&2.222(16)&1.803(8)&1.4968(28)&1.2553(13)&1.0303(20)\\
6&3.88(6)&2.845(35)&2.220(12)&1.8045(45)&1.5025(25)&1.2672(12)&1.0517(12)\\
8&3.88(12)&2.835(65)&2.234(22)&1.8065(85)&1.5020(40)&1.2722(22)&1.0650(20)\\
10&3.90(10)&2.885(45)&2.230(12)&1.8035(75)&1.5045(30)&1.2744(26)&1.0650(20)\\
12&3.88(4)&2.83(5)&2.220(14)&1.8035(65)&1.5038(25)&1.2759(19)&1.0823(18)\\
14&3.88(4)&2.862(42)&2.225(14)&1.8055(35)&1.5050(25)&1.2764(14)&1.0880(20)\\
16&3.89(7)&2.845(30)&2.225(15)&1.803(7)&1.5047(19)&1.2768(18)&1.0932(12)\\
$T_{\rm N}(p)$&3.2308&2.3604&1.8205&1.4427&1.1530&0.9102&0.6792\\
\br
\end{tabular}
\end{indented}
\end{table}

 We have no definite principle to estimate
the value of $T_{0}(p)$ which is the extrapolated value
of $T_{0}(L,p)$ to $L\rightarrow \infty$.  
No clear size dependence, however, can be seen at $p=0.65-0.8$. Thus, we perform naive extrapolation to $L\rightarrow \infty$ in this region.
It can be seen that the value of $T_{0}(L,p)$  slightly increases
as $L$ increases at $p=0.85-0.95$.  We have found  that
the extrapolation by the $N^{-1}$-law works fairly well at $p=0.85$ and 
$p=0.9$,
where $N$ is the total number of the spins of the system.
Figure 3 shows a plot of $T_{0}(L,p)$ versus $1/N$ at $p=0.9$, where we have
 estimated
that $T_{0}(p=0.9)=1.278(3)$.
We have also found that the extrapolation by the $L^{-1/2}$-law works fairly 
well at $p=0.95$.
The estimated values of $T_{0}(p)$ for various $p$  are shown in table 2.
Strictly speaking, we cannot justify the above extrapolaton at each $p$.
We can say, however, that the extrapolated values, $T_{0}(p)$, might not change
drastically even if we use other extrapolation laws, 
and there seems to be no possibility 
that the value of $T_{0}(p)$ becomes smaller than the temperature
of the Nishimori line at each $p$.

\begin{table}
\caption{The estimated value of $T_{0}(p)$ for various $p$. $T_{\rm N}(p)$ 
is the
temperature of the Nishimori line at $p$.}
\begin{indented}
\item[]\begin{tabular}{@{}lll}
\br
$p$&$T_{0}(p)$&$T_{\rm N}(p)$\\
\mr
0.65&3.88(6)&3.2308\\
0.7&2.845(65)&2.3604\\
0.75&2.217(19)&1.8205\\
0.8&1.805(9)&1.4427\\
0.85&1.506(3)&1.1530\\
0.9&1.278(3)&0.9102\\
0.95&1.154(3)&0.6792\\
\br
\end{tabular}
\end{indented}
\end{table}

Figure 4 shows the estimated values of $T_{0}(p)$ for various $p$ in the $p-T$
 plane, which are denoted by black squares.  The dashed line
denotes the Nishimori line,
 and the open circles denote the 
paramagnetic-ferromagnetic phase boundary above the multicritical point[19].
We can see that the region where $C_{2}(p,T)$ takes the negative value
is fairly large. We can also see that the region mentioned
above contains the Nishimori line.
There are several numerical calculations about the 
paramagnetic-ferromagnetic phase boundary of the two-dimensional
$\pm J$ Ising model[13,19-22,27]. 
It is noted that all the calculations are
almost consistent with each other and they show that, above
 the multicritical point,
the concentration, $p$, of the phase boundary increases as the temperature
increases.
Thus, a part of the paramagnetic-ferromagnetic phase boundary
is included in the region where $C_{2}(p,T)$ takes the negative value.
In that region, the specific heat cannot diverge.
Namely, the specific heat does not diverge at least on
a part of the 
paramgnetic-ferromagnetic phase boundary
above the multicritical point.  

\section{Monte Carlo simulations near the paramagnetic-ferromagnetic
phase boundary}

In order to elucidate  the region  where $C_{2}(p,T)$ takes the negative value
near the 
paramagnetic-ferromagnetic phase
boundary, we have performed the Monte Carlo simulations for more larger
lattices, $L=31-121$, at $p=0.88-0.92$. 
In the calculations,  we take the skew boundary condition in one direction
and the periodic boundary condition in the other direction.

The condition of the simulation is shown in table 3.
M.C.S. denotes the Monte Carlo step  of the simulation, which is chosen to be $20\tau$, where the relaxation time, 
$\tau$, is evaluated  by the statistical time-independent method[23], and $N_{b}$ is
the number of the bond configurations.

\begin{table}
\caption{The condition of the Monte Carlo simulation.}
\begin{indented}
\item[]\begin{tabular}{@{}llll}
\br
$p$&$L$&${\rm M.C.S.}$&$N_{b}$\\
\mr
0.88&31&400000&240\\
&61&800000&96\\
&121&800000&120\\
0.89&31&600000&240\\
&61&5000000&120\\
&121&16000000&80\\
0.9&31&800000&320\\
&61&1400000&160\\
&91&30000000&64\\
0.91&31&400000&480\\
&61&10000000&96\\
&121&20000000&32\\
0.92&31&600000&720\\
&61&1400000&120\\
&121&800000&64\\
\br
\end{tabular}
\end{indented}
\end{table}

The temperature dependence of  $C_{2}(p,T)$ near the zero point  for $L=91$ 
at $p=0.9$ is shown in figure 5, where we have estimated that
$T_{0}(L=91,p=0.9)=1.2765(45)$.  The estimated
values of $T_{0}(L,p)$ for various $L$ and $p$
are shown in table 4.

\begin{table}
\caption{The estimated values of $T_{0}(L,p)$ for various $L$ and $p$.}
\begin{indented}
\item[]\begin{tabular}{@{}lllll}
\br
$p$&$L=31$&$L=64$&$L=96$&$L=121$\\
\mr
0.88&1.3485(55)&1.3535(45)&&1.3555(35)\\
0.89&1.3075(45)&1.303(5)&&1.3055(55)\\
0.9&1.286(4)&1.2775(45)&1.2765(45)&\\
0.91&1.269(4)&1.2665(45)&&1.2685(45)\\
0.92&1.249(4)&1.2485(45)&&1.2485(35)\\

\br
\end{tabular}
\end{indented}
\end{table}

Though $T_{0}(L,p)$ takes similar value even if the lattice size,
$L$, changes, we have  extrapolated  the values of $T_{0}(p,L)$ to $L\rightarrow \infty$ by the $N^{-1}$-law.
We show the estimated values of $T_{0}(p)$ for various $p$ in table 5 with
the temperature of the Nishimori line, $T_{\rm N}(p)$.

\begin{table}
\caption{The values of $T_{0}(p)$ for various $p$. $T_{\rm N}(p)$ 
is the
temperature of the Nishimori line at $p$.}
\begin{indented}
\item[]\begin{tabular}{@{}lll}
\br
$p$&$T_{0}(p)$&$T_{\rm N}(p)$\\
\mr
0.88&1.355(5)&1.0038\\
0.89&1.3045(55)&0.9566\\
0.9&1.275(5)&0.9102\\
0.91&1.267(5)&0.8644\\
0.92&1.2485(55)&0.8188\\
\br
\end{tabular}
\end{indented}
\end{table}

The results are also shown in figure 6. 
The black circles and squares denote $T_{0}(p)$ 
evaluated  by the Monte Carlo simulations and by the numerical transfer matrix method, respectively.
The dashed line denotes the Nishimori line, and the open circles denote the paramagnetic-ferromagnetic phase boundary above the multicritical point[16].
It can be seen that both
 results by the numerical transfer matrix method and Monte Carlo
simulations are consistent with each other.

The crossing
 point of the paramagnetic-ferromagnetic phase boundary and the boundary of the region where
$C_{2}(p,T)$ takes the negative value  has been estimated to be 0.8985(15).
Therefore, we  conclude that the specific heat cannot diverge on the
paramagnetic-ferromagnetic phase boundary  at least for
$p_{\rm c}\leq p \leq 0.8985(15)$, where $p_{\rm c}$ is the 
concentration of the
multicritical point. There are many numerical estimates of $p_{\rm c}$:
0.8905(5)[14],0.8872(8)[24],0.886(3)[25], 0.8906[26]  and 0.8907(2)[27]. 
Recently,
there is a conjecture about the exact value of $p_{\rm c}$, which insists
that $p_{\rm c}=0.889972$[28].

\section{Conclusions}
We have investigated the property of the specific heat 
of the two-dimensional $\pm J$ Ising model by
the numerical transfer matrix method and Monte Carlo simulations
from a new point of view.
We have estimated the region where $C_{2}(p,T)$ takes the negative value
in the $p-T$ plane, which is one of the characteritic of the frustrated system, and reflects the non-trivial degeneracy of the ground state.
By the numerical transfer matrix method for $L=4-16$, 
and the Monte Carlo simulations for $L=31-121$, the region 
have been found to be fairly large in the $p-T$ plane, which
includes the Nishimori-line.  

Moreover, the region includes a part of
the paramagnetic-ferromagnetic phase boundary, on which the specific
heat cannot diverge. 
Thus, our results indicate that the specific heat cannot diverge 
on the paramagnetic-ferromagnetic phase boundary at least near and above 
the multicritical point.
On the other hand, there are several literatures
that, near the pure ferromagnetic point($p\geq 0.92$), the specific heat diverges at most
logarithmically[13,14]. 
Both results are not directly in conflict with each other,
since the range of the concentration of the calculations does not overlap.
It is natural, however, to think that the nature of  the 
phase transition does not change on the paramagnetic-ferromagnetic
phase boundary above the multicriticsl point.
In the literatures which insist the divergence of the specific heat,
the change of the peak height of the specific heat of various lattice size
was investigated[13,14]. It is a very subtle problem whether the specific heat 
diverges or
remains finite, when the divergence is so weak as the logarithmic divergence.
Our extrapolation, however, is straightforward and the values of $T_{0}(p)$
might not change drastically even when there exist other system-size corrections,
and there seems to be
 no possibility that the value of $T_{0}(p)$ becomes smaller than
the temperature of the Nishimori line at each $p$.
Thus,
we insist
that  the specific heat  of the two-dimensional
$\pm J$ Ising model  does not diverge 
on the whole range of the paramagnetic-ferromagnetic phase boundary
above the multicritical point,
though we cannot exclude the possibility that a singular point
exists on the paramagnetic-ferromagnetic phase boundary,
which separates the nature of the specific heat.

\ack
The authors would thank Dr. Y. Ozeki and Dr. K. Hukushima for their
useful discussions.
The calculations were made on HITAC SR8000 at University of Tokyo.
and at the Institue for
Solid State Physics in University of Tokyo.

\begin{figure}[htbp]
\begin{center}
\includegraphics[width=65mm]{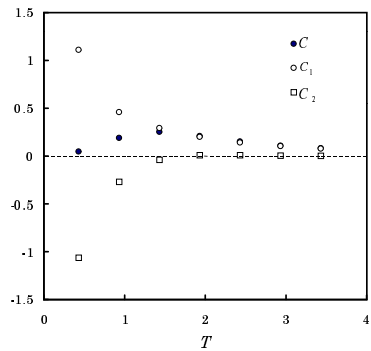}
\end{center}
\caption{The temperature dependence of $C(p,T)$, $C_{1}(p,T)$
and $C_{2}(p,T)$ for $L=4$ at $p=0.8$.}
\end{figure}

\begin{figure}[htbp]
\begin{center}
\includegraphics[width=65mm]{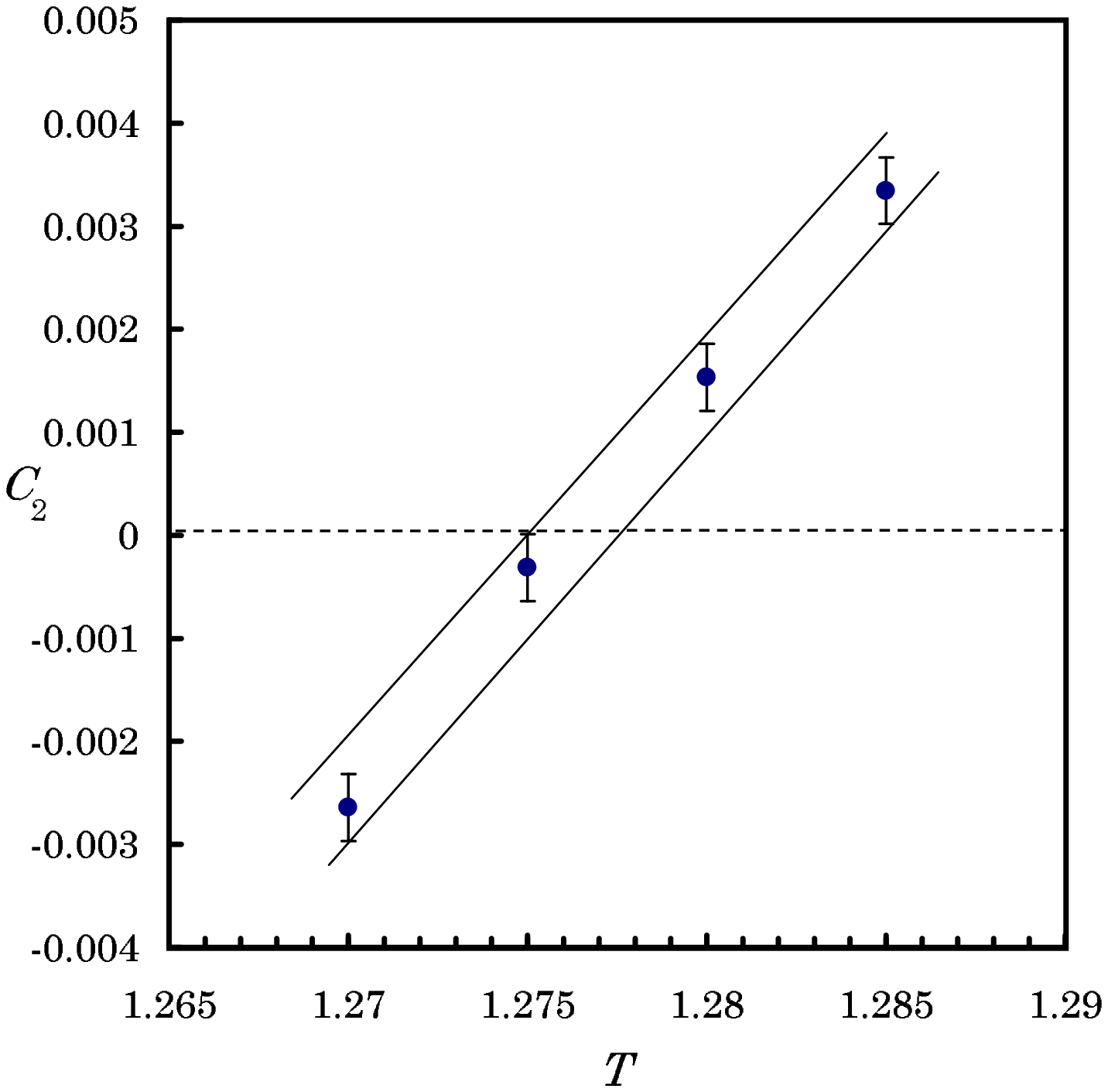}
\end{center}
\caption{The temperature dependence of $C_{2}(p,T)$ near the zero point
for $L=14$ at $p=0.9$.}
\end{figure}

\begin{figure}[htbp]
\begin{center}
\includegraphics[width=65mm]{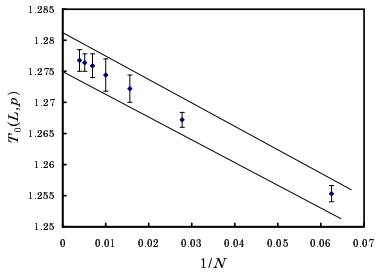}
\end{center}
\caption{A plot of $T_{0}(L,P)$ versus $1/N$ at $p=0.9$.}
\end{figure}

\begin{figure}[htbp]
\begin{center}
\includegraphics[width=65mm]{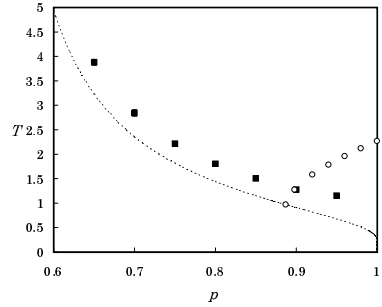}
\end{center}
\caption{The estimated values of $T_{0}(p)$ for various $p$ in the $p-T$ plane,
which are denoted by black squares.
The dashed line denotes the Nishimori line, and the open circles denote the paramagnetic-ferromagnetic 
phase boundary above the multicritical point.}
\end{figure}

\begin{figure}[htbp]
\begin{center}
\includegraphics[width=65mm]{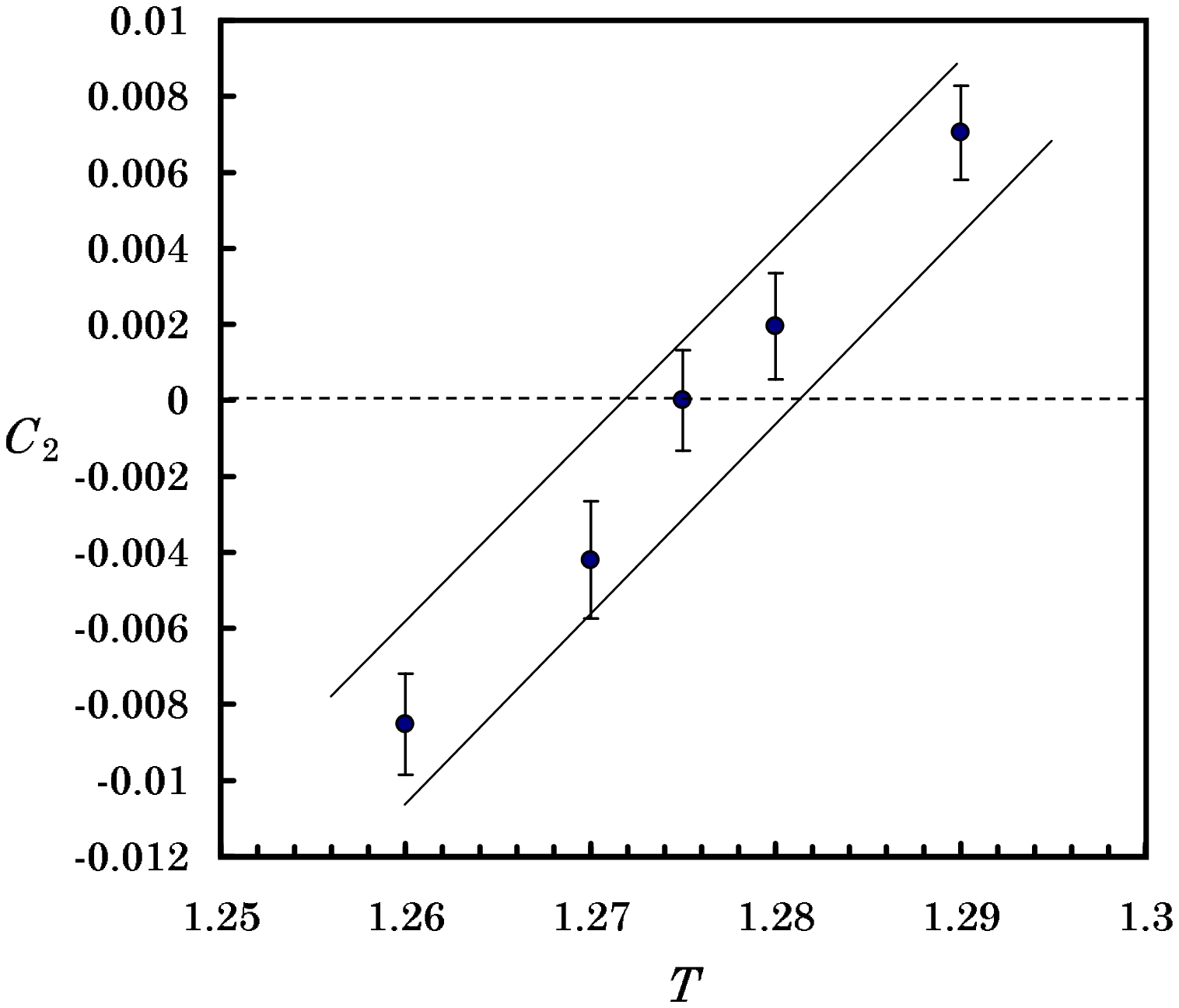}
\end{center}
\caption{The temperature dependence of $C_{2}(p,T)$ near the zero point  for $L=91$ at $p=0.9$.}
\end{figure}

\begin{figure}[htbp]
\begin{center}
\includegraphics[width=65mm]{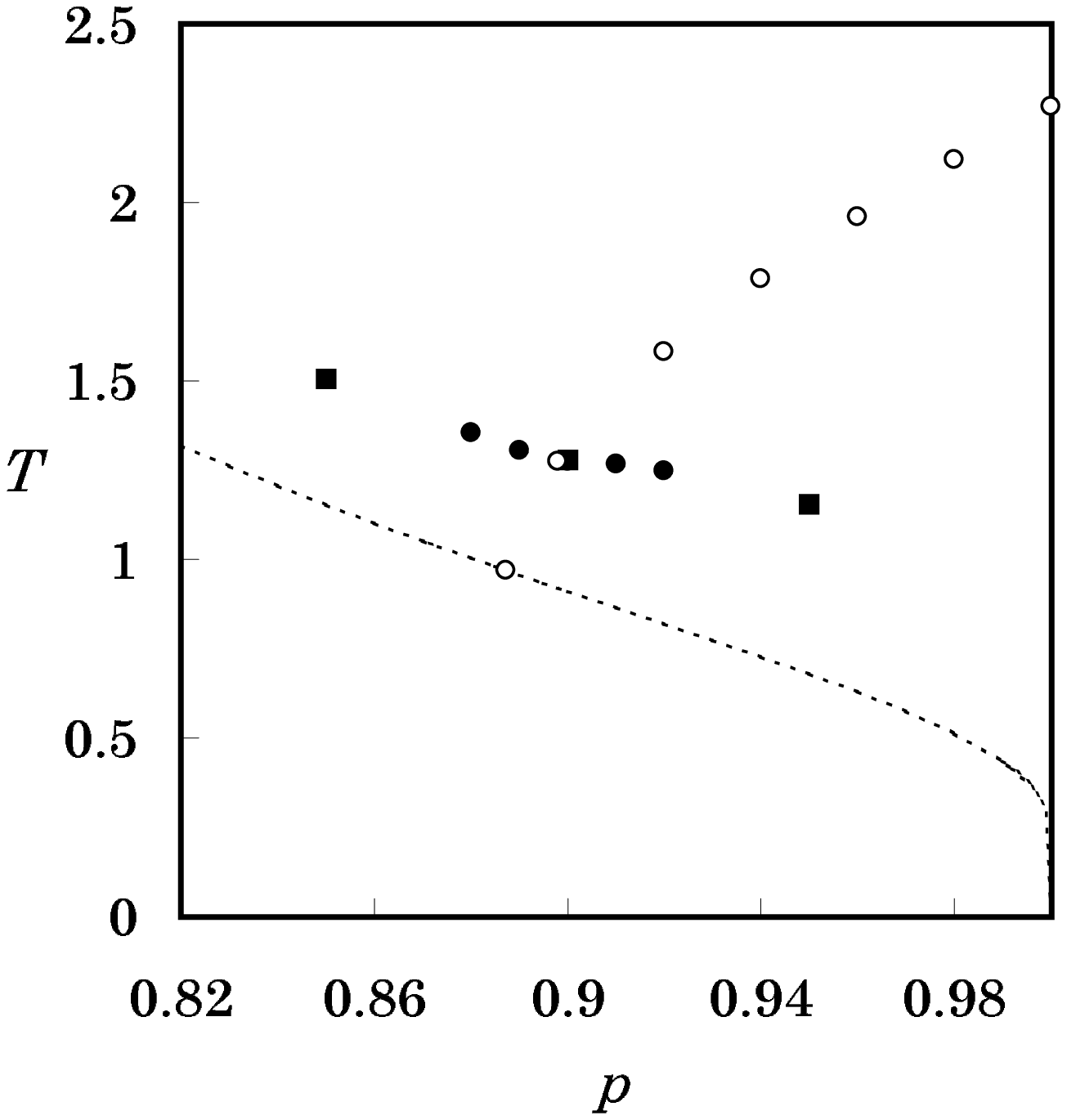}
\end{center}
\caption{The estimated values of $T_{0}(p)$ for various $p$ in the $p-T$ plane.
The black circles and squares denote the values of $T_{0}(p)$ by the Monte Carlo simulations and by the numerical transfer matrix method, respectively.
The dashed line denotes the Nishimori line, and the open circles denote the paramagnetic-ferromagnetic phase boundary above the multicritical point.}
\end{figure}


\section*{References}
\begin{thebibliography}{100}
\bibitem{Harris} Harris A B 1974
{\it J. Phys. C}  {\bf 7} 1671
\bibitem{Dotsenko} Dotsenko B V S and Dotsenko V S 1983
{\it Advan. Phys.} {\bf 32} 129
\bibitem{Shanker} Shanker R 1987
{\it Phys. Rev. Lett.} {\bf 23} 2466.
\bibitem{Blackman} Blackman J A and Poulter J 1984
{\it J. Phys. C} {\bf 17} 107
\bibitem{Wang} Wang J -S, Selke W and Dotsenko V S 1990
{\it Physica A} {\bf 164} 221
\bibitem{Wiseman} Wiseman S and Domany E 1995
{\it Phys. Rev. B} {\bf 51} 3074
\bibitem{Stauffer} Stauffer D, Reis F D A A, de Queiroz S L A and
dos Santos R R 1997
{\it Int. J. Mod. Phys. C} {\bf 8} 1209
\bibitem{Selke} Selke W, Shchur L N and Vasilyev O A 1998
{\it Physica A} {\bf 259} 388
\bibitem{Ballesteros} Ballesteros H G, Fernandez L A, Mayor V M,
Sudupe A M, Parisi G and Ruiz-Lorenzo J J 1997
{\it J. Phys. A} {\bf 30} 8379
\bibitem{Tamaribuchi} Tamaribuchi T and Takano F 1980
{\it Prog. Theor. Phys.} {\bf 64} 1212
\bibitem{Kim} Kim J and Patrascioiu A 1994
{\it Pys. Rev. Lett.} {\bf 72} 2785
\bibitem{Ziegler} Zielger K 1991
{\it Europhys. Lett.} {\bf 14} 415
\bibitem{Inoue} Inoue M 1995
{\it J. Phys. Soc. Jpn.} {\bf 64} 3699
\bibitem{Reis} Aarao Reis F D A, de Queiroz S L A and
dos Santos R R 1998
{\it Phys. Rev. B} {\bf 60} 6740
\bibitem{Nishimori} Nishimori H 1981
{\it Prog. Theor. Phys.} {\bf 66} 1169
\bibitem{Jacobson}Jacobsen J L  and Picco M 2002
{\it Phys. Rev. E} {\bf 65} 026113
\bibitem{Honecker} Honecker A, Picco M and Pujol P 2001
{\it Phys. Rev. Lett.} {\bf 87} 047201
\bibitem{Hukushima} Hukushima K 2000
{\it J. Phys. Soc. Jpn.} {\bf 69} 631
\bibitem{Ito} Ito N, Ozeki Y and Kitatani H 1999
{\it J. Phys. Soc. Jpn.} {\bf 68} 803
\bibitem{Ozeki} Ozeki Y and Nishimori H 1987
{\it J. Phys. Soc. Jpn} {\bf 56} 3265
\bibitem{Kitatani} Kitatani H and Oguchi T 1990
{\it J. Phys. Soc. Jpn.} {\bf 59} 3823
\bibitem{Kitatani2} Kitatani H and Oguchi T 1992
{\it J. Phys. Soc. Jpn} {\bf 61 } 1598
\bibitem{Kikuchi} Kikuchi M and Ito N 1993
{\it J. Phys. Soc. Jpn.} {\bf 62} 3052 
\bibitem{Ozeki2} Ozeki Y and Ito N 1998
{\it J. Phys. A} {\bf 31} 5451
\bibitem{Singh} Singh R R P and Adler J 1996
{\it Phys. Rev. B} {\bf 54} 364
\bibitem{Honecker} Honecker A, Picco M and Pujol P 2001
{\it Pjys. Rev. Lett.} {\bf 87} 047201.
\bibitem{Merz} Merz F and Chalker J T 2002
{\it Phys. Rev. B} {\bf 65} 054425.
\bibitem{Nishimori2} Nishimori H and Nemoto K 2002
{\ J. Phys. Soc. Jpn.} {\bf 71} 1198

\end{thebibliography}
\end{document}